\documentclass[aps,amssymb,amsmath,twocolumn,secnumarabic]{revtex4}
\usepackage{amsfonts}
\usepackage{dsfont}
\usepackage{pdfpages}
 \usepackage{verbatim}
\usepackage{tikz}

\usepackage{xcolor}

\usepackage[utf8]{inputenc}
\usepackage[OT1]{fontenc}
\usepackage[english]{babel}

\usepackage{amsmath,amssymb,amsfonts,mathrsfs}
\usepackage[amsmath,thmmarks]{ntheorem}

\usepackage{graphicx}
\usepackage{soul}
\usepackage{pdfpages}
\usepackage[colorlinks,bookmarks=false,linkcolor=blue,urlcolor=blue,
citecolor=magenta]{hyperref}
\usepackage{physics}

\usepackage{stmaryrd}

\usepackage{varioref}
\usepackage{mathtools}
\usepackage[h]{esvect}
\usepackage{array}
\usepackage{listings}
\usepackage{booktabs}
\usepackage[mathscr]{euscript}
\usepackage{amsbsy}
\usepackage{bbm}

\usepackage[justification=centering]{caption}

\usepackage{lmodern}

\begin{document}

\title{From Gaudin Integrable Models to $d$-dimensional Multipoint Conformal Blocks}

\author{Ilija Buri\' c$^a$, Sylvain Lacroix$^b$, Jeremy A. Mann$^a$, Lorenzo Quintavalle$^a$
and Volker Schomerus$^{a}$}

\affiliation{$^a$DESY Theory Group, DESY Hamburg, Notkestrasse 85, D-22603 Hamburg,\\
$^b$ II. Institut f\"ur Theoretische Physik, Universit\"at Hamburg, Luruper Chaussee 149, D-22761 Hamburg\\
Zentrum für Mathematische Physik, Universität Hamburg, Bundesstrasse 55, D-20146 Hamburg}

\date{September, 2020}

\begin{abstract}
In this work we initiate an integrability-based approach to multipoint conformal blocks for higher dimensional
conformal field theories. Our main observation is that conformal blocks for $N$-point functions
may be considered as eigenfunctions of integrable Gaudin Hamiltonians. This provides us
with a complete set of differential equations that can be used to evaluate multipoint blocks.

\end{abstract}

\maketitle

\vspace*{-7cm}
\noindent
{\tt DESY 20-157\\
\tt ZMP-HH/20-25\\
\tt SAGEX-20-22-E}\\[4.2cm]

\section{Introduction}

Conformal quantum field theories (CFTs) play an important role for our
understanding of phase transitions, quantum field theory and even the
quantum physics of gravity, through Maldacena’s celebrated holographic
duality. Since they are often strongly coupled, however, they are very
difficult to access with traditional perturbative methods. Polyakov's
famous conformal bootstrap program provides a powerful non-perturbative
handle that allows to calculate critical exponents and other dynamical
observables using only general features such as (conformal) symmetry,
locality and unitarity \cite{Polyakov:1974gs}. The program has had
impressive success in $d=2$ dimensions \cite{Belavin:1984vu} where it
produced numerous exact solutions. During the last decade, the bootstrap
has seen a remarkable revival in higher dimensional theories
with new numerical as well as analytical incarnations. This has produced
many stunning new insights, see e.g.\ \cite{Poland_2019} for a review and
references, including record precision computations of critical exponents
in the critical 3D Ising model \cite{Kos:2016ysd,Simmons-Duffin:2016wlq}.
Despite these advances, it is evident that significant further developments
are needed to make these techniques more widely applicable, beyond a few
special theories.

One promising avenue would be to study bootstrap consistency conditions for
$N$-point correlators with $N > 4$ fields. Note that the success in $d=2$ is
ultimately based on the ability to analyze correlation functions
with any number of stress tensor insertions. But the extension of the bootstrap
constraints in $d > 2$ beyond 4-point functions has been hampered by very
significant technical problems, see \cite{Rosenhaus:2018zqn, Parikh:2019ygo,
Fortin:2019dnq,Goncalves:2019znr,Parikh:2019dvm,Fortin:2019zkm,Irges:2020lgp,Fortin:2020yjz,
Fortin:2020ncr,Zhou:2020ptb,Pal:2020dqf,Fortin:2020bfq,Hoback:2020pgj} for
recent publications. To overcome these challenges is the main goal of our
work.

The central tool for CFTs in general and for the conformal bootstrap in particular
are conformal partial wave expansions. These were introduced in \cite{Ferrara:1973vz}
to separate correlation functions into kinematically determined conformal blocks (partial
waves) \footnote{In this letter we shall not distinguish between the two notions and simply
use the term conformal block.} and expansion coefficients which contain all the dynamical
information. For 4-point correlators, the relevant blocks are now well understood in any
$d$, though only after some significant effort. Here we shall lay the foundations for a
systematic extension to multipoint (MP) blocks. Our approach extends a remarkable observation
in \cite{Isachenkov:2016gim} about a relation between 4-point blocks and exactly solvable
(integrable) Schroedinger problems.

To understand the key challenge in developing a theory of MP
conformal blocks, let us consider a 5-point function of scalar
fields. In more than two dimensions one can build five independent conformally
invariant cross ratios from $N=5$ points. Correlation functions can be evaluated
through repeated use of Wilson's operator product expansion (OPE). We
may picture this process with the help of an OPE diagram, such as the
one shown in Fig. 1. For $N=5$ points, any such diagram contains two
intermediate fields. The scaling weights $\Delta$ and spins $l$ of
these intermediate fields provide four quantum numbers. This is not
sufficient to resolve the dependence of the $5$-point function on the
five cross ratios. The missing fifth quantum number is somehow associated
with the choice of so-called tensor structures at the vertices of an OPE
diagram. In the case of the 5-point function in $d > 2$, the middle vertex
in Fig. 1 gives rise to one additional quantum number. But what precisely
is the nature of this quantum number and how can it be measured?
\footnote{Note that this question has not been addressed
in any of the recent papers on MP blocks \cite{Rosenhaus:2018zqn,Parikh:2019ygo,
Fortin:2019dnq,Goncalves:2019znr,Parikh:2019dvm, Fortin:2019zkm,Irges:2020lgp,
Fortin:2020yjz,Fortin:2020ncr,Zhou:2020ptb,Anous:2020vtw,Pal:2020dqf,Fortin:2020bfq,
Hoback:2020pgj,Fortin:2020zxw}.}
\begin{figure}[thb]
\centering
\captionsetup{labelfont=bf,width=.9\linewidth}
\includegraphics[scale=1,clip]{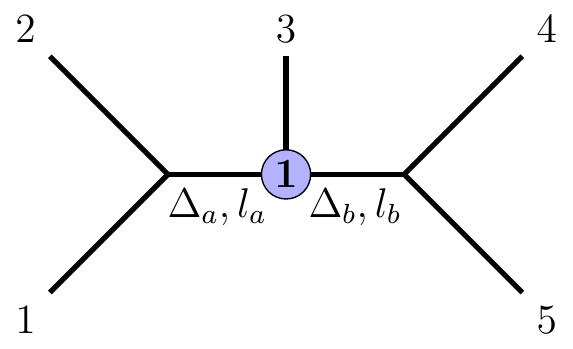}
\caption{\small OPE diagram for a 5-point function. The corresponding 5-point
conformal block depends on five quantum numbers which are measured by
four Casimir operators and one new vertex DO.}
\label{figure1}
\end{figure}

In order to describe our answer let us turn to the most basic
description of conformal blocks, the so-called shadow formalism
\cite{Ferrara:1972uq}. The latter provides integral formulas for
conformal blocks that are reminiscent of Feynman integrals.
Finding analytical expressions in terms of special functions or
even just efficient numerical evaluations requires significant
technology. One crucial tool in the theory of Feynman integrals
is to consider them as solutions of some differential equations. In
their important work, Dolan and Osborn followed this same strategy
and characterized shadow integrals as eigenfunctions of a set of
Casimir differential operators (DOs) \cite{Dolan:2003hv}. By
studying these differential equations they were able to harvest
decisive new results on the conformal blocks \cite{Dolan:2003hv,
Dolan:2011dv}.

Shadow integral representations for MP blocks are also known. In
order to evaluate these, one may want to follow very much the same strategy
that was used for 4-point functions. It is indeed relatively straightforward
to write down MP generalizations of the Casimir operators of \cite{Dolan:2003hv}.
In the case of 5-point functions in $d>2$ there are four of them. Their eigenvalues
measure the weight and spin of the intermediate fields. But as we explained above,
this is not sufficient. We need one more DO that commutes with the four Casimir
operators to measure a fifth quantum number. This appears to set the stage for
some integrable system and indeed, as we shall show below, the four Casimir
operators along with the fifth missing one can be constructed as
commuting Hamiltonians of the famous Gaudin integrable model
\cite{gaudin1976diagonalisation,Gaudin_book83}, in a certain
limit. The statement may be established more generally but the 5-point
function of scalar fields is the first case for which we have worked out
these DOs explicitly.

Let us now outline the content of this short note. In the next section we
review how to construct shadow integral representations for MP
functions with a particular focus on the choice of tensor structures at the
vertices. We introduce a novel basis of 3-point tensor structures that
enables us to characterize the shadow integral, and hence the blocks, as
common eigenfunctions of a set of five commuting DOs.
In section 3, we explain how these operators can be constructed systematically from
Hamiltonians of the Gaudin integrable model by taking a special limit. Four
of the five DOs are Casimir operators while the fifth one
measures the choice of tensor structure. We conclude with an outlook on our
forthcoming paper \cite{Buric:2020}, extensions and applications to the
higher dimensional conformal bootstrap.

\section{Multipoint Shadow Integrals}\label{sec:shadow}

In order to state our results precisely, we shall briefly review some basics
of the shadow integral formalism. The shadow formalism turns the graphical
representation of a conformal block, such as that of Fig. 1, into an
integral formula. Just as in the case of Feynman integrals, the `shadow
integrand' is built from relatively simple building blocks that are assigned
to the links and 3-point vertices in the associated OPE diagram. For a scalar
5-point function, the most complicated vertex contains one scalar leg and two
that are carrying symmetric traceless tensor (STT) representations. In order
to write this vertex, we shall employ polarization spinors $z \in \mathbb{C}^d$
(see \cite{Dobrev:1976aa,Dobrev:1977qv,Costa:2011aa,Costa:2011ab}) to convert
spinning operators in STT representations into objects of the form
\begin{equation}
\mathcal{O}_{\Delta,l} (x;z) = \mathcal{O}_{\Delta,l}^{\nu_1...\nu_l} (x)
z_{\nu_1}...z_{\nu_l} \equiv \mathcal{O}^{\underline{\nu}}_{\Delta,l}(x)
z_{\underline{\nu}}\ .
\end{equation}
The usual contraction of the STTs can be re-expressed as an integral over
$\mathbb{C}^d$ as follows \cite{Bargmann_1977}
\begin{equation}
\mathcal{O}^{\underline{\nu}} (x)
\mathcal{O}'_{\underline{\nu}} (x') = \!
\int_{\mathbb{C}^d}\!\! \dd^{2d} z\, \delta(z^2) \rho(\bar{z} \cdot z)
\,\mathcal{O}(x;\bar{z}) \mathcal{O}' (x';z),
\end{equation}
\begin{equation}
\rho(t)  = \left( \frac{2}{\pi} \right)^{d-1} \frac{(16 t)^{1-d/4}}
{\Gamma (d/2-1)} K_{(d/2-2)} ( 2 \sqrt{t}),
\end{equation}
where $\mathcal{O}$ and $\mathcal{O}'$ are fields of equal spin and
$K$ is the modified Bessel function of the second kind. In building shadow
integrands, the function $\rho$ plays a role analogous to the propagator in
Feynman integrals. Having now converted field multiplets into functions, the
3-point vertex with one scalar leg and two STT legs takes the form
\begin{align}
\Phi^t_{acb}(x;z) & =
\langle \mathcal{O}_{\Delta_a,l_a}(x_a;z_a)    \mathcal{O}_{\Delta_c}(x_c)
\mathcal{O}_{\Delta_b,l_b} (x_b;z_b)  \rangle = \nonumber \\[2mm]
& \hspace*{-1cm} \frac{ (X_{bc;a} \cdot z_a)^{l_a}
(X_{ca;b} \cdot z_b)^{l_b}}{(X_{ab;c}^2)^{\frac{-\Delta_c}{2}}
(X_{ca;b}^2)^{\frac{l_b-\Delta_b}{2}} (X_{bc;a}^2)^{\frac{l_a-\Delta_a}{2}}}
t\left( X \right),\label{eq:3point}
\end{align}
if $l_a-l_b \in 2\mathbb{Z}$ and vanishes otherwise. Here we have used the standard notation
\begin{equation}
X_{ij;k}^{\mu} := \frac{x_{ik}^{\mu}}{x_{ik}^2}- \frac{x_{jk}^{\mu}}{x_{jk}^2}
= - X_{ji;k}^{\mu},  \,\,\, X_{ij;k}^2 = \frac{x_{ij}^2}{x_{ik}^2 x_{jk}^2},
\end{equation}
with $x_{ij} = x_i-x_j$, and we have dubbed $X$ the unique independent cross-ratio that can
be constructed from $(x_a,x_b,x_c;z_a,z_b)$,
\begin{equation}
X = \frac{1}{2 x_{ab}^4} \frac{z_{a\mu} \left( x_{ab}^2 \delta^{\mu\nu} - 2 x_{ab}^{\mu}
x_{ab}^{\nu} \right) z_{b\nu}}{(z_a \cdot X_{bc;a}) (z_b \cdot X_{ca;b})} \ .
\end{equation}
To a large extent, the function $t(X)$ that appears in the 3-point vertex is left
undetermined by conformal symmetry. The only constraints come from the action of
the $SO(d-1)$ subgroup  that stabilizes three points in $\mathbb{R}^d$, as well as
the parity operator in $\mathrm{O}(d)$. For parity-even vertices, the function
$t(X)$ belongs to the space $W^+_t$ of polynomials of order at most $\textit{min}
(l_a,l_b)$. Parity-odd vertices with a single scalar leg only exist in $d=3$. In
this case, the function $t(X) \in W_t^-$ must be chosen such that $t(X)/\sqrt{X(1-X)}$
is a polynomial of order at most $\textit{min}(l_a,l_b)-1$. In total, the admissible
functions $t(X)$ span a vector space of dimension
\begin{equation}
n_{ab} = \sum_\pm \textit{dim}\,W_t^\pm  = \left\{
\begin{array}{ll}  2\textit{min}(l_a,l_b)+1,  & \quad d = 3, \\
\textit{min}(l_a,l_b)+1, & \quad d > 3.
\end{array} \right .
\end{equation}
The integer $n_{ab}$ counts  the number of 3-point tensor structures
\cite{Costa:2011ab}. Note that $n_{ab} = 1$ if either $l_a=0$ or $l_b=0$ which
means that $t$ is a constant factor if there are two or three scalar legs. We
shall therefore simply drop the corresponding vertex factors $t$ when using
formula \eqref{eq:3point} for vertices with two scalar legs.

Having described the vertex, we can now write down (shadow) integrals for
any desired $N$-point function in the so-called comb channel, in which every
OPE includes at least one of the external scalar fields. For $N=5$ external
scalar fields of weight $\Delta_i, i=1, \dots,5$ the shadow integrals read
\begin{align}
& \Psi^{(\Delta_1,...,\Delta_5)}_{(\Delta_a,\Delta_b;l_a,l_b;t)} (x_1,...,x_5) =
\label{eq:5pshadow}\\
  & = \prod_{s =a,b} \int_{\mathbb{R}^{d}} \hspace*{-2mm} \dd^d x_s
 \int_{\mathbb{C}^d} \hspace*{-2mm}\dd^{2d} z_s
 \delta(z_s^2) \rho(\bar{z}_s\cdot z_{s}) \Phi_{12\tilde{a}}(x_1,x_2,x_a;\bar{z}_a)
 \nonumber \\
&  \hspace*{1.5cm} \times \Phi^t_{a3b}(x_a,x_3,x_b;z_a,z_b)
\Phi_{\tilde{b}45}(x_b,x_4,x_5;\bar{z}_b) \ . \nonumber
\end{align}
Here the tilde on the indices of the first and third vertex means that we use eq.\
\eqref{eq:3point} for two scalar legs but with $\Delta_a$ and $\Delta_b$ replaced by
$d-\Delta_a$ and $d-\Delta_b$, respectively.

After splitting off some factor $\Omega$ that accounts for the nontrivial covariance
law of the scalar fields under conformal transformations,
\begin{align}
&\Psi^{(\Delta_i)}_{(\Delta_a,\Delta_b;l_a,l_b;t)} (x_i)  =
\Omega^{(\Delta_i)} (x_i)
\psi^{(\Delta_{12},\Delta_3,\Delta_{45})}_{(\Delta_a,\Delta_b;l_a,l_b;t)}
(u_1,...,u_5),\nonumber  \\[2mm]
& \Omega^{(\Delta_i)} (x_i)  :=
(X_{23;1}^2)^{\frac{\Delta_1}{2}} \,
\prod_{i=2}^4 (X_{i+1,i-1;i}^2)^{\frac{\Delta_i}{2}} \,
(X_{34;5}^2)^{\frac{\Delta_5}{2}}, \nonumber
\end{align}
with $\Delta_{ij} = \Delta_i-\Delta_j$ as usual, the shadow integral \eqref{eq:5pshadow}
gives rise to a finite conformal integral that defines the conformal block $\psi$ as a
function of five conformally invariant cross ratios $u_i$. These integrals depend on the
choice of $(\Delta_a,l_a)$, $(\Delta_b,l_b)$ and the function $t(X)$. Our goal is to
compute this uninviting looking integral.

The strategy we have sketched in the introduction is to write down five
differential equations for these blocks. Four of these are given by the
eigenvalue equations for the second and fourth order Casimir operators
for the intermediate channels,
\begin{equation} \label{eq:Casimireq}
\mathcal{D}^s_p \psi^{(\Delta_{12},\Delta_3,\Delta_{45})}_{(\Delta_a,\Delta_b;l_a,l_b;t)}(u)
= C_p^s \psi^{(\Delta_{12},\Delta_3,\Delta_{45})}_{(\Delta_a,\Delta_b;l_a,l_b;t)}\ ,
\end{equation}
where $p=2,4$ and $C_p^s$ denotes the eigenvalue of the $p$-th order Casimir
operator in the representation $(\Delta_s,l_s)$ for $s=a,b$. The explicit
form of the DOs $\mathcal{D}^s_p$ can be worked out and the resulting
expressions resemble those in \cite{Dolan:2003hv}.

But we are missing one more differential equation which we shall construct in the
next section. It will turn out that shadow integrals are  eigenfunctions of a
fifth DO provided we prepare a very
special basis $t_n(X), n=1, \dots, n_{ab},$ in the space of 3-point tensor
structures. We can characterize these functions $t_n(X)$ as eigenfunctions
of a particular fourth order DO
\begin{equation}\label{eq:Vertop}
H^{(d,\Delta_i,l_i)} = h_0(X)+\sum_{q=1}^4 h_{q}(X)
 X^{q-1}(1-X)^{q-1}\partial_X^{q}\ ,
\end{equation}
where $h_q = h_q^{(d,\Delta_i,l_i)} $ are polynomials of order at most three,
see Supplemental Material at [URL] for concrete expressions.  The operator
$H$, which has several remarkable properties, appears to be new. For our discussion
it is most important to note that $H$ leaves the two subspaces $W_t^\pm$ invariant
whenever both $l_a$ and $l_b$ are integer. Consequently, it specifies a
special basis $t_n$ of functions $t(X)$ in the space of tensor structures,
\begin{equation} \label{eq:Httaut}
H^{(d,\Delta_i,l_i)} t_n(X) = \tau_n t_n(X)\ ,  \quad  n = 0, \dots, n_{ab}\ .
\end{equation}
Explicit formulas for the eigenvalues $\tau_n$ and the eigenfunctions
$t_n(X)$ can be worked out, and it is this basis of 3-point tensor structures
that we will use to write down differential equations for the associated shadow
integrals.

\section{Multipoint Blocks and Gaudin Hamiltonians}
\label{sec:gaudin}
\def\a{\alpha}
\def\H{\mathcal{H}}
\def\Ht{\widetilde{\mathcal{H}}}

Our goal now is to characterize the shadow integrals through a complete
set of five differential equations. These will take the form of eigenvalue
equations for a set of commuting Gaudin Hamiltonians. In order to state
precise formulas we need a bit of background on Gaudin models~\cite{gaudin1976diagonalisation,Gaudin_book83}. Let us
begin with a central object, the so-called Lax matrix,
\begin{equation}
 \mathcal{L}(w)  = \sum_{i=1}^N \frac{\mathcal{T}^{(i)}_\a T^\a}{w-w_i}
 = \mathcal{L}_\a(w) T^\a\ .
\label{eq:Lax}
\end{equation}
Here, $w_i$ are a set of complex numbers, $T_\a$ denotes a basis of
generators of the conformal Lie algebra in $d$ dimensions and $T^\a$ its dual basis with
respect to an invariant bilinear form. The object $\mathcal{T}^{(i)}_\a$ is
the standard first order DO that describes the behavior of
a scalar primary field $\mathcal{O}(x_i)$ of weight $\Delta_i$ under the
conformal transformation generated by $T_\a$.

Given some conformally invariant symmetric tensor $\kappa_p$ of degree $p$ one
can construct a family $\mathcal{H}_p(w)$  of commuting operators
as~\cite{Feigin:1994in,Talalaev:2004qi,Molev:2013}
\begin{equation}\label{eq:HpGaudin}
\H_p(w) = \kappa^{\a_1 \cdots \a_p}_p \mathcal{L}_{\a_1}(w) \cdots
\mathcal{L}_{\a_p}(w) + \dots  \ ,
\end{equation}
where the dots represent correction terms expressible as lower
degree combinations of the Lax matrix components $\mathcal{L}_\a(w)$ and
their derivatives with respect to $w$. For $p=2$ such correction terms are
absent. The correction terms are necessary to ensure that the families
commute,
\begin{equation}\label{eq:com}
[\, \H_p(w) \, , \, \H_q(w') \, ] = 0\ ,
\end{equation}
for all $p,q$ and all $w,w' \in \mathbb{C}$. In the case where $d\geq 3$, the conformal
algebra possesses two independent invariant tensors of second and fourth degree \footnote{For $d>3$, there are also additional invariant tensors. However, we will not need those for what follows.}.
We therefore obtain two families of commuting DOs that act on
functions of the coordinates $x_i$.

It is a well-known fact that these families commute with the diagonal action of
the conformal algebra, i.e.
\begin{equation}
[\, \mathcal{T}_\a\,  ,\,  \H_p(w)\, ] = 0 \quad \textit{ where }
\ \mathcal{T}_\a = \sum_{i=1}^N \mathcal{T}_\a^{(i)}\ .
\end{equation}
Hence the commuting families $\H_p(w)$ of operators descend
to DOs on functions $\psi(u)$ of the conformally
invariant cross ratios $u$.
\smallskip

The functions $\H_p(w)$ provide several continuous families of
commuting operators. Only a finite set of these operators are
independent. There are many ways of constructing such sets of
independent operators, e.g. by taking residues of $\H_p(w)$ at
the singular points to give just one example. For the moment,
any such set still contains $N$ parameters $w_i, i= 1, \dots,
N$. Without loss of generality we can set three of these complex
numbers to some specific value, e.g. $w_1 = 0, w_{N-1}=1, w_{N}
= \infty$ so that we remain with $N-3$ complex parameters our
Gaudin Hamiltonians depend on.

Now we adapt the Gaudin model to the study
of MP blocks. In the latter context we insist that the set of
commuting operators we work with allows us to measure the weights $\Delta$
and spins $l$ of  fields that are exchanged in intermediate channels,
as do the MP Casimir operators. So, in
order for the Gaudin Hamiltonians to be of any use to us, we must
ensure that they include all such Casimir operators. For this to be
the case, we are forced to make a very special choice of the remaining
parameters $w_r$ and to consider specific limits of these parameters
\footnote{Such limits have also been considered in~\cite{Chervov:2007dn,Chervov:2009}
to study bending flow Hamiltonians and their generalizations~\cite{kapovich1995,
kapovich1996,FlaschkaMillson,Falqui_2003}}. Let us explain this here for $N=5$.
Setting $w_2 = \varpi^2$ and $w_3 = \varpi$ we can define
\begin{equation}\label{eq:limit}
\Ht_p(w) := \lim_{\varpi \rightarrow 0}
\varpi^p \H_p(\varpi w) \quad , \ p = 2,4.
\end{equation}
The new functions $\Ht_p$ take values in the space of $p^{\textit{th}}$ order
DOs on cross ratios. They possess singularities at three
points only, namely at $w = 0,1, \infty$. Let us note that taking the limit
$\varpi \rightarrow 0$ does not spoil commutativity of these Hamiltonians.

After performing the special limit on the parameters $w_r$ we can now
extract the MP Casimir operators rather easily. In fact, it is
not difficult to check that
\begin{equation}\label{eq:caslim}
\mathcal{D}^a_p = \lim_{w \rightarrow 0} w^p \Ht_p(w) \quad , \quad
\mathcal{D}^b_p = \lim_{w \rightarrow \infty} w^p \Ht_p(w)
\end{equation}
for $p=2,4$. Any additional independent operator we can obtain from
$\Ht_p(w)$ may be used to measure a fifth quantum number. One
can show that the two second order Casimir operators $\mathcal{D}^s_2, s=a,b$
exhaust all the independent operators that can be obtained from
$\Ht_2(w)$. The family $\Ht_4(w)$, on the other hand, indeed supplies one
independent operator in addition to the fourth order Casimir operators $\mathcal{D}^s_4,
s=a,b$. We propose to use the operator $\mathcal{V}_4$ defined through
\begin{equation} \label{eq:vertexop}
\Ht_4\left( w= 1/2 \right) = 16\,\mathcal{V}_4  + \dots \, ,
\end{equation}
where the dots represent quadratic terms coming from the corrections in eq. \eqref{eq:HpGaudin}.
In the particular limit $\varpi\rightarrow 0$ that we consider here, these corrections can be
reexpressed in terms of the quadratic Casimirs $\mathcal{D}^s_2, s=a,b$, and can thus be discarded
without spoiling commutativity of $\mathcal{V}_4$ with the Casimirs. An explicit computation then
shows that $\mathcal{V}_4$ is expressed in terms of the conformal generators $\mathcal{T}^{(i)}_\alpha$ as
\begin{equation}
\mathcal{V}_4 = \kappa^{\alpha_1\cdots\alpha_4}_4 \mathcal{S}_{\alpha_1} \cdots \mathcal{S}_{\alpha_4},
\qquad \mathcal{S}_\alpha = \mathcal{T}_\alpha^{(1)} + \mathcal{T}_\alpha^{(2)} - \mathcal{T}_\alpha^{(3)}.
\end{equation}
The explicit form of $\mathcal{V}_4$ as a DO acting on functions $\psi(u)$
of five cross ratios will be spelled out in our forthcoming publication \cite{Buric:2020}.
Our central claim is that the 5-point shadow integrals $\psi$ we discussed
in the previous subsection are joint eigenfunctions of the four Casimir
operators, see eq.\ \eqref{eq:Casimireq}, and of the vertex operator we
defined through eq.\ \eqref{eq:vertexop},
\begin{equation} \label{eq:V4ev}
 \mathcal{V}_4\
 \psi^{(\Delta_{12},\Delta_3,\Delta_{45})}_{(\Delta_a,\Delta_b;l_a,l_b;t_n)}(u)
 = \tau_n \,
 \psi^{(\Delta_{12},\Delta_3,\Delta_{45})}_{(\Delta_a,\Delta_b;l_a,l_b;t_n)}(u)
\ ,
\end{equation}
where the eigenvalues $\tau_n$ coincide with those that appeared in
eq.\ \eqref{eq:Httaut} when describing the particular choice of a basis
$t_n(X)$ of tensor structures. These five differential equations characterize
the shadow integral completely.
\medskip

Before we conclude, let us briefly sketch how the above exposition extends to the comb
channel of $N$-point functions in arbitrary dimension $d$. In this case, the
Lax matrix \eqref{eq:Lax} of the Gaudin model depends on $N$ complex parameters
$w_i$. We can set three of these to the values $w_1=0$, $w_{N-1}=1$ and $w_N=
\infty$ before scaling the remaining ones as $w_i=\varpi^{N-i-1}, i=2, \dots,
N-2$ in terms of a single complex parameter $\varpi$ that we send to zero.
Generalizing our construction of the commuting families of operators in eq.\ \eqref{eq:limit}
we now introduce
\begin{equation} \label{eq:Hpr}
\Ht_p^{[r]}(w) := \lim_{\varpi \rightarrow 0}
\varpi^{(N-r-2)p} \H_p(\varpi^{N-r-2} w),
\end{equation}
where $p = 2,4, \dots$ enumerates the different (Casimir) invariants of the $d$-dimensional
conformal algebra and $w \in \mathbb{C}$ is the spectral parameter. Through the label $r\in
\lbrace 1, \dots,N-2\rbrace$ we characterize different ways to perform the scaling limit
of the original Gaudin Hamiltonians. It is not difficult to show that the resulting family
of commuting Hamiltonians includes all the Casimir operators that are needed to measure the
weight and spin of intermediate fields, similarly to eq.\ \eqref{eq:caslim}. The other
Hamiltonians extracted from the families \eqref{eq:Hpr} then provide additional commuting
operators characterizing the vertices in the $N$-point conformal block (note that the range of our index $r$ indeed allows us to enumerate these vertices). One thereby expects to complete the full set of Casimir operators into a system of
independent commuting operators that suffices to characterize the dependence of $N$-point comb
channel blocks on all conformal cross ratios, for arbitrary dimension $d$ and arbitrary
choice of representations for external fields. We have checked this claim for various
choices of $N$ and $d$.

For $d=3$, an $N$-point function with scalar external fields involves $3N-10$ cross ratios. The intermediate fields in the
comb channel OPE diagram are characterised by $2N-6$ Casimir operators, of degree two and four. In
addition, each of the $N-4$ internal vertices is associated with an operator $\mathcal{V}_4^{[r]}$, extracted similarly to $\mathcal{V}_4$ in eq. \eqref{eq:vertexop} as
\begin{equation} \label{eq:vertexopr}
\Ht^{[r]}_4\left( w = 1/2 \right) = 16\,\mathcal{V}^{[r]}_4  + \dots \, ,
\end{equation}
where $r\in\lbrace 2,\dots,N-3 \rbrace$ \footnote{Note that for the case of scalar external fields, the extremal vertices of the comb channel diagram are trivial, which is why we restrict $r$ to the range $\lbrace 2,\dots,N-3 \rbrace$ in this case.}. The spectrum of these $N-4$ operators is independent of
$r$ and is still given by the eigenvalues $\tau_n$ we introduced in section \ref{sec:shadow}.
With the additional index $r\in\lbrace 2,\dots,N-3 \rbrace$ on the left hand side of the vertex eigenvalue equation
\eqref{eq:V4ev}, we obtain enough differential equations to characterize 3-dimensional $N$-point
blocks in the comb channel.

\section{Conclusions and Outlook}

In this work we initiated a systematic construction of MP
conformal blocks in $d \geq 3$. Our advance relies on a characterization
of MP conformal blocks as wave functions of Gaudin integrable
models, which extends a similar relation between 4-point blocks and
integrable Calogero-Sutherland models uncovered in \cite{Isachenkov:2016gim}.
More specifically, we have explained that for a very special choice of tensor
structures at the 3-vertices $\Phi$ in the shadow integrand of eq.\
\eqref{eq:5pshadow}, the corresponding shadow integral becomes a joint
eigenfunction of a complete set of commuting DOs.
The latter are Hamiltonians of special limits of the Gaudin model.

While we have explained the main ideas within the example of $5$-point
functions, the strategy and in particular the relation with Gaudin models
is completely general, i.e. it extends to $N > 5$ and even spinning external
operators, with appropriate changes (see for instance the end of section \ref{sec:gaudin} for the comb channel case). Starting from six points, there exist topologically
distinct channels that can include vertices in which all three legs carry spin, such
as the so-called snowflake channel for $N=6$ \cite{Fortin:2020yjz}. Such vertices involve
functions $t$ of several variables and hence the choice of basis in the
space of tensor structures needs to be extended. As we increase the
dimensions $d$, links can carry new representations beyond STT. Treating
more generic links only requires us to consider higher order Casimir
operators. Through the relation to Calogero-Sutherland models
\cite{Isachenkov:2016gim}, their solution theory is well known, see e.g.\
\cite{Isachenkov:2017qgn}. In this sense, links do not pose a significant
new complication for the construction of MP blocks in any $d$.

In forthcoming work \cite{Buric:2020} we will explain in detail how to
construct the vertex DOs, both for the shadow integrand
and the shadow integral, and we shall spell out explicit formulas for
all five DOs that characterize the shadow integrals for $5$-point
functions. This can then serve as a starting point to evaluate 5-point
blocks explicitly, e.g.\ through series expansions or Zamolodchikov-like
recursion formulas, similar to those used for 4-point blocks
\cite{Dolan:2011dv,Hogervorst:2013sma,Kos:2013tga,Kos:2014bka,
Penedones:2015aga,Isachenkov:2017qgn}.

Obviously, it would be very interesting to extend these constructions
of DOs to $6$-point blocks, to develop an evaluation
theory and to initiate a MP bootstrap for $d>2$. As we have
argued in the introduction, taking bootstrap constraints from MP
correlation functions seems like a good strategy. Key examples for
initial studies include the $O(n)$ Wilson-Fisher fixed points with
$n=2,3$ that describe the $\lambda$-point in Helium or the ferromagnetic
phase transition, respectively. The current state-of-the-art for $n=2$
was set recently in \cite{Chester_2020,Liu:2020tpf}, using $4$-point
mixed correlator and analytic bootstrap. Since $6$-point functions of
a single scalar field contain the information of infinitely many
mixed $4$-point functions, the MP bootstrap for $N=6$ can be
expected to provide significantly stronger bounds.

While we were completing this letter Vieira et al. issued the
paper \cite{Goncalves:2019znr} in which they initiate a MP light-cone
bootstrap. With the techniques we propose here, it should be possible
to study light-cone blocks along with systematic corrections
in the vicinity of the strict light-cone limit and for any desired
channel. We will come back to these topics in future work.
\smallskip

\noindent
{\bf Acknowledgements:} We are grateful to Gleb Arutyunov, Aleix Gimenez-Grau,
Mikhail Isachenkov, Madalena Lemos, Pedro Liendo, Junchen Rong, Joerg
Teschner and Beno\^it Vicedo for useful discussions. This project
received funding from the German Research Foundation DFG under
Germany’s Excellence Strategy – EXC 2121 ,,Quantum Universe'' –
390833306 and from the European Union’s Horizon 2020 research and
innovation programme under the MSC grant agreement No. 764850 “SAGEX”.


\appendix
\vspace{-.501 cm}
\section{The Vertex Operator $H$}\label{app:H}
Since it might be of interest for some readers we list all the coefficients $h_q(X)$ of the Hamiltonian \eqref{eq:Vertop} that is used to define our basis $t_n$ of 3-point tensor
structures. Except for a constant term in $h_0$ which depends a bit on the
precise choice of the fifth Gaudin Hamiltonian we extract, all coefficients
are symmetric w.r.t. exchange of $a$ and $b$. Hence we will split them as
\begin{equation*}
    h^{(\Delta_a,l_a;\Delta_c;\Delta_b,l_b)}(X) =\chi^{(\Delta_a,l_a;\Delta_c;\Delta_b,l_b)}(X) +\ a \leftrightarrow b
\end{equation*}
and display the polynomials $\chi(X)$ instead of $h(X)$. Despite its relevance for representation theory, we have not found the fourth
order operator \eqref{eq:Vertop} in the existing literature on orthogonal
polynomials, except for some special cases.

\begin{eqnarray*}
	\chi_4 & = &  8\,,\\[2mm]
	\chi_3 & = & 32 X \left(l_a -2\right)-4 \left(4 l_a+ 2\Delta _c-d-8\right)\,,\\[2mm]
	\chi_2 & = & 16 X^2 \Bigl(l_a^2+2 l_a l_b -9 l_a+7\Bigr)\\
	& & -4 X \Bigl(4 l_a^2+8 l_a l_b+2 l_a \left(2 \Delta _c-d-18\right)\\
	& & \quad +  2\Delta _a \Delta _b-2d \Delta _a -(4+d) \Delta _c+d^2+2d+28\Bigr)\\
	& & +2 \Bigl((l_a+l_b)^2+2l_a\left(2\Delta _c-2d-4\right)\\
	& &	\quad -2\Delta _a^2 +\Delta _c^2 + 2\Delta _a \Delta _b-2(d+2) \Delta _c+6d+4\Bigr)\,,
	\end{eqnarray*}
\vspace{-.7cm}
\onecolumngrid
\begin{eqnarray*}
	\chi_1 & = & 16 X^3 \left(l_a-1\right) \left(l_b-1\right) \left(l_a+l_b-2\right)\\
	& & -2 X^2 \Bigl(24 l_a^2 (l_b-1) + 2 l_a l_b (2\Delta _c-24-d) + (4l_a-2) ( 2\Delta _a \Delta _b-d(\Delta _a+\Delta_b+ \Delta _c)+18+d^2) +12\Bigr)\\
	& & +2 X \Bigl(2l_a^2 \left(4 l_b-d-2\right)+4l_a l_b \left(\Delta _c-d-3\right)+ 2 l_a  \bigl(4 \Delta _a \Delta _b -2d( \Delta _a+\Delta_b + \Delta _c-3) +d^2+4\bigr)\\
	& & \qquad \quad +(d-2)(2\Delta _a^2-\Delta _c^2)-4 \Delta _a \Delta _b -2d(d-4) \Delta _a+d^2 \Delta _c
	-8d\Bigr)\\
	& & +(d-2)\Bigl((l_a+l_b)^2+ 4l_a (\Delta _c-2) -2\Delta _a^2+\Delta _c^2+2 \Delta _a \Delta _b-4 \Delta _c+4\Bigr)\,,\\[2mm]
	\chi_0 & = &-8 X^2\,  l_a\left(l_a-1\right)  l_b \left(l_b-1\right) +4 X l_a l_b \Bigl( 2 l_al_b-4 l_a + 2 \Delta _a \Delta _b-2d \Delta _a -(d-2)\Delta _c+d^2-d+2\Bigr) + \textit{const}\,.
\end{eqnarray*}

\twocolumngrid

\end{document}